\documentclass[sigconf]{acmart}

\usepackage{booktabs} % For formal tables

\setcopyright{rightsretained}
\usepackage{graphicx}
\usepackage{graphics}
\usepackage{scalefnt}
\usepackage[ruled,vlined]{algorithm2e}
\usepackage{tikz}
\usetikzlibrary{trees,snakes}
\usepackage{tikz-qtree}
\usepackage{multirow}
\usepackage{hhline}
\usepackage{url}
\usepackage{caption}
\usepackage{subfig}
\usepackage{amssymb,amsmath}
\usepackage{fancyvrb}
\usepackage{textcomp}
\usepackage{color, colortbl}
\definecolor{orange}{rgb}{1,0.5,0}
\newcommand{\commentcode}[2]{\hspace{-0.0001px}{\scalefont{#1}\textcolor{gray}{\textit{#2}}}}
\usepackage{enumitem}
\setlist[itemize]{leftmargin=*}

\acmYear{}
\copyrightyear{}

\begin{document}
\title{XTreePath: A generalization of XPath to handle real world structural variation}
%\titlenote{Produces the permission block, and
%  copyright information}
%\subtitle{Extended Abstract}
%\subtitlenote{The full version of the author's guide is available as
%  \texttt{acmart.pdf} document}

\author{Joseph Paul Cohen}
%\authornote{Dr.~Trovato insisted his name be first.}
\affiliation{%
  \institution{University of Massachusetts Boston}
  %\streetaddress{P.O. Box 1212}
  \city{Boston} 
  \state{MA} 
}
\email{joecohen@cs.umb.edu}

\author{Wei Ding}
%\authornote{The secretary disavows any knowledge of this author's actions.}
\affiliation{%
  \institution{University of Massachusetts Boston}
  %\streetaddress{P.O. Box 1212}
  \city{Boston} 
  \state{MA} 
}
\email{ding@cs.umb.edu}

\author{Abraham Bagherjeiran}
%\authornote{This author is the
%  one who did all the really hard work.}
\affiliation{%
  \institution{ThinkersR.Us}
  %\streetaddress{1 Th{\o}rv{\"a}ld Circle}
  \city{Sunnyvale} 
  \state{CA} 
  }
\email{abagher@thinkersr.us}

% \affiliation{%
% \affiliation{%
% \affiliation{%
%   \country{China}
% }

% \affiliation{%


% \author{Julius P.~Kumquat}

\renewcommand{\shortauthors}{Cohen, J.P. et al.}

\begin{abstract}

We discuss a key problem in information extraction which deals with wrapper failures due to changing content templates. A good proportion of wrapper failures are due to HTML templates changing to cause wrappers to become incompatible after element inclusion or removal in a DOM (Tree representation of HTML). We perform a large-scale empirical analyses of the causes of shift and mathematically quantify the levels of domain difficulty based on entropy. We propose the XTreePath annotation method to captures contextual node information from the training DOM. We then utilize this annotation in a supervised manner at test time with our proposed Recursive Tree Matching method which locates nodes most similar in context recursively using the tree edit distance. The search is based on a heuristic function that takes into account the similarity of a tree compared to the structure that was present in the training data. We evaluate XTreePath using 117,422 pages from 75 diverse websites in 8 vertical markets. Our XTreePath method consistently outperforms XPath and a current commercial system in terms of successful extractions in a blackbox test. We make our code and datasets publicly available online. 
\end{abstract}


%\keywords{ACM proceedings, \LaTeX, text tagging}

\maketitle

\section{Introduction}

We address a key problem in information extraction which deals with wrapper failures due to changing content templates. A wrapper is best described by Kushmerick as a ``procedure, specific to a single information resource, that translates a [webpage] query response to relational form'' \cite{kushmerick_wrapper_1997}. Wrappers are required because much of the desired data on the Internet is presented using HTML templates instead of well formed (XML or JSON) data or unstructured freeform text \cite{flesca_web_2004}. Extracting data, such as stock, flight, and product information, from websites that use HTML templates is difficult because wrapper methods have difficulty dealing with changes to HTML structure. 

We believe a good proportion of wrapper failures are due to HTML templates changing and cause wrappers to become incompatible after the element inclusion or removal of DOM (Tree representation of HTML). For example, an additional ``On Sale'' element is included above a product. This \textit{shift} \cite{dalvi_robust_2009, zhai_web_2005} may require manual retraining of wrappers which is a burden to users. Empirically we find over 50\% of our web sample contains shift at various levels and we have a detailed discussion about shift in Section \S \ref{sec:shiftanalysis}. 

To handle the problem of shift we need to take in more information from the DOM \cite{zhai_web_2005, gulhane_web-scale_2011, Kosala2002, Chidlovskii2003, Omari2017, dalvi_automatic_2011}. Instead of only extracting statistics from the training data we extract an entire sub-tree structure, like Reis \cite{reis_automatic_2004}, and use this during wrapper induction to create a supervised information extraction method.

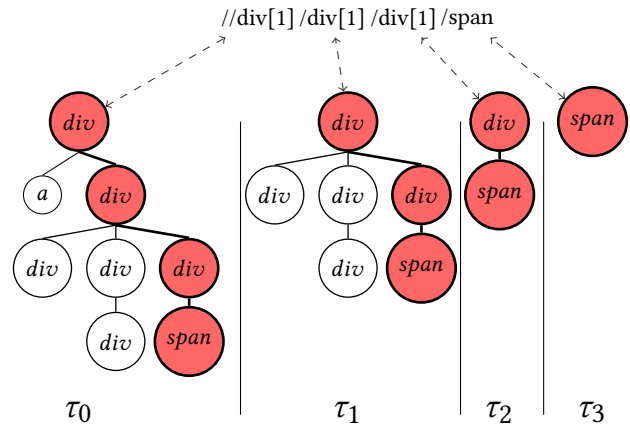
\begin{figure}
\begin{center}

\begin{tikzpicture}[scale=0.65,
leaf/.style={circle, draw=none, fill=black,
        text centered, text=white,},
redfill/.style={fill=red!60},
bold/.style={line width=1.0pt,redfill},
notbold/.style={line width=0.5pt}
]

%\draw (2.1,0) -- ++(0,-5);
\begin{scope}[xshift=3.2cm]
	\node[bold] [circle,draw,bold] (1){$div$}
% 			<a>a</a>
			child {node [circle,draw] (ao) {$a$}}	
		child[bold] {node [circle,draw,bold] (10o) {$div$}
% 				<div id="4"></div>
				child[black, notbold] {node [circle,draw] (4o) {$div$}}
% 				<div id="9">
				child[black, notbold] {node [circle,draw] (9o) {$div$}
% 					<div id="2">div2</div>
					child {node [circle,draw] (2o) {$div$}}
				}
% 				</div>
% 				<div id="5">
				child[bold] {node [circle,draw,bold] (5o) {$div$}
% 					<span id="6">span</span>
					child[bold] {node [circle,draw,bold] (span1o) {$span$}}
				}
% 				</div>
			};
% 			</div>
\node [yshift=-110pt,scale=1.5] (a0){$\tau_0$};
\end{scope}
\draw (6.5,0) -- ++(0,-6);
\begin{scope}[xshift=+8.7cm]
\node [circle,draw, bold] (10){$div$}

% 				<div id="4"></div>
			child[black, notbold] {node [circle,draw] (4o) {$div$}}
% 			</div>
% 				<div id="9">
			child[black, notbold] {node [circle,draw] (9o) {$div$}
% 					<div id="2">div2</div>
				child {node [circle,draw] (2o) {$div$}}
			}
% 				</div>
% 				<div id="5">
			child[bold] {node [circle,draw,bold] (5o) {$div$}
% 					<span id="6">span</span>
				child[bold] {node [circle,draw,bold] (span1) {$span$}}
			};
% 				</div>
\node [yshift=-110pt,scale=1.5] (a1){$\tau_1$};
\end{scope}

\draw (11,0) -- ++(0,-6);
\begin{scope}[xshift=+11.8cm]
\node [circle,draw,bold] (5){$div$}
% 					<div id="2">div2</div>
		child[bold] {node [circle,draw,bold] (2) {$span$}};
\node [yshift=-110pt,scale=1.5] (a2){$\tau_2$};
\end{scope}

\draw (12.7,0) -- ++(0,-6);
\begin{scope}[xshift=+13.7cm]
\node [circle,draw,bold] (span1){$span$};
\node [yshift=-110pt,scale=1.5] (a3){$\tau_3$};
\end{scope}

\begin{scope}[yshift=+60pt,xshift=+180pt]
\node [xshift=+10pt] (1bx) {//div[1]}; 
\node [xshift=+38pt] (10bx) {/div[1]}; 
\node [xshift=+65pt] (5bx) {/div[1]}; 
\node [xshift=+89pt] (span1bx) {/span}; 
\end{scope}

% tree to xpath before
\draw[<->,dashed,color=black!80] (1) -- node[above,sloped] {} (1bx); 
\draw[<->,dashed,color=black!80] (10) -- node[above,sloped] {} (10bx); 
\draw[<->,dashed,color=black!80] (5) -- node[above,sloped] {} (5bx); 
\draw[<->,dashed,color=black!80] (span1) -- node[above,sloped] {} (span1bx); 

\end{tikzpicture}
\end{center}
\caption{
Here we present the view of a tree path in relation to an XPath. The red nodes show the nesting of trees within each other in the tree path. This is comparable to the XPath $//div[1]/div[1]/div[1]/span$.
}
\label{fig:TPath}
\end{figure}

Figure \ref{fig:TPath} gives an example of XTreePath. We start by finding the least common ancestor of the nodes of interest (product name, price, etc) and extract a relative XPath as well as the nodes of the DOM for every step of that XPath. We then use this XTreePath with our Recursive Tree Matching method which performs a heuristic graph search in the target DOM to find the most similar sub-trees to those in the training data. Because we are matching sub-trees we can handle large horizontal and vertical shifts as long as some unique traits about the DOM are preserved. 

Furthermore, XTreePath is compatible by design, which complements XPath but does not replace it. We only utilize the recursive tree matching lookup process after an XPath has failed, which minimizes runtime. This allows an XPath-based method to adopt XTreePath without sacrificing existing speed or quality. Existing research has discovered many methods to construct robust XPaths a priori. Our presented method can exist in parallel with these methods as they are continually advanced in order to achieve better overall accuracy.

\noindent Thus, our main contributions are as follows: 
\begin{itemize}
\setlength\itemsep{0em}
	\item {\bf Method:} We propose the XTreePath method which is a generalization of XPaths where tree structure is also stored and used during wrapper induction.
	\item {\bf Algorithm Design:} We propose a Recursive Tree Matching method and a dynamic programming solution to perform wrapper induction.
	\item {\bf Theoretical Analysis:} We formally define and analyze the problem of shift theoretically and empirically. We mathematically quantify the levels of domain difficulty based on entropy and study a large representative dataset. 
    \item {\bf Robustness:} We evaluate XTreePath using 117,422 pages from 75 diverse websites in 8 vertical markets. 
    \item {\bf Reproducibility:} Our code and data are open-sourced at \url{http://kdl.cs.umb.edu/w/datasets/}.
    
\end{itemize}

\section{Related Work}
\label{sec:relatedwork}

In the field of information extraction there are two primary categories of annotation learning: supervised and unsupervised approaches. We take a supervised approach and combine positional (XPath \cite{Anton2005}) and ontological (Tree Structure) concepts \cite{kayed_survey_2006, Nasti2016, kushmerick_wrapper_2000} together.

In Dalvi and Parameswaran \cite{dalvi_robust_2009, dalvi_automatic_2011, parameswaran_optimal_2011} focused on supervised annotation learning algorithms tolerant to noise in the training data. They enumerate many XPath wrappers using a probabilistic ranking system to pick the best one. The output of these methods is XPath annotations which differentiates our method as XTreePath would complement this method and not replace it.

The context of nodes has been used to create annotations themselves but not during extraction as our method does. In 2009 Zheng \cite{zheng_efficient_2009} and Fang \cite{Fang2017} used a ``broom'' structure inside the HTML DOM to represent both records and generated wrappers. This work was motivated to capture lists of products instead of creating wrappers tolerant to shift.

Tree similarity has been used in unsupervised information extraction approaches to find common sub-trees in websites by Zhai in 2005 \cite{zhai_web_2005} and used as a method to locate lists from web pages by Jindal in 2010 \cite{jindal_generalized_2010}. These methods focus on locating interesting data but not in a way of imposing a label as our method XTreePath does. 

A break away from just carrying XPaths forward from the training was discussed by Omari \cite{Omari2017}. In this work they would learn a decision decision tree to predict which XPath fragments should be used at test time.

Moving beyond XPath is not a new concept. There has been work on tree automata induction by \cite{Kosala2002} which aims to learn a deterministic finite automata which will process the DOM tree and accept nodes which contain the information of interest. The mechanics of this are very different but we can also argue that a DFA does not take the neighborhood of the tree into account when accepting a node.

Possibly the most similar approach to our method was proposed by Reis\cite{reis_automatic_2004}. Here they use the tree edit distance to determine if a webpage contains a specific type of content (news) based on tree structure. We take this concept and structure it into a generalized XPath representation that can be used for supervised information extraction.

\section{Shift Analysis}
\label{sec:shiftanalysis}

A \textbf{shift} of a web page occurs when a modification of the page causes the inclusion, removal, or substitution of DOM elements which changes the DOM tree representation. Not all shifts are bad. A shift only becomes a problem when it causes wrappers to become incompatible and return no result or an incorrect result. Here compatibility is defined as whether the DOM structure of a wrapper matches the DOM structure of a web page.

\newcommand*\circled[1]{\tikz[baseline=(char.base)]{
            \node[shape=rectangle,red,draw,inner sep=1pt] (char) {#1};}}

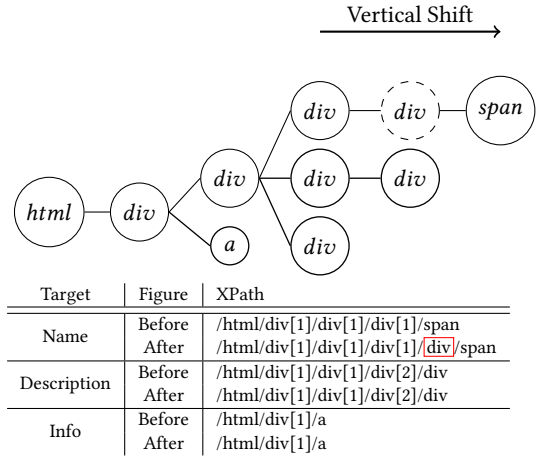
\begin{figure}[h]

%\begin{center}

\hspace{10pt}
%{\scalefont{0.8}
	\begin{tikzpicture}[scale=0.8,
	leaf/.style={circle, draw=none, fill=black,
	        text centered, text=white},grow=right, %
	%redfill/.style={fill=red!60},
	bold/.style={},
	notbold/.style={line width=0.5pt},
	sibling distance=32pt	
	]
	%\draw (6.4,0) -- ++(0,-5);
	\begin{scope}
	\node[bold]  [circle,draw] (fa){$html$}
	%	\node [circle,draw] (f){$div_1$}
		child[bold]  {node [circle,draw, bold] (1a) {$div$}
			child[black, notbold]  {node [circle,draw] (aa) {$a$}}
			child[bold]  {node [circle,draw, bold] (10a) {$div$}
	% 				<div id="4"></div>
					child[black, notbold]  {node [circle,draw] (4a) {$div$}}
	% 				<div id="9">
					child[black, notbold]  {node [circle,draw] (9a) {$div$}
	% 					<div id="2">div2</div>
						child {node [circle,draw] (2a) {$div$}}
					}
	% 				</div>
	% 				<div id="5">
					child[bold]  {node [circle,draw,bold] (5a) {$div$}
						child[bold]  {node [circle,draw,dashed, bold] (3a) {$div$}
		% 					<span id="6">span</span>
							child[bold]  {node [circle,draw, bold] (span1a) {$span$}}
						}
					}
	% 				</div>
				}
	% 			</div>
	% 			<a>a</a>
			};
			
	\draw[thick,->] (4.5,3) --  node[above] {Vertical Shift} (7.5,3);

	\end{scope}
	
	%\draw[<->,dashed,color=gray!100] (f) -- node[above,sloped] {} (fa);
	
	\end{tikzpicture}
%	}
%}
\begin{center}
{\scalefont{0.8}
\begin{tabular}{c|c|l}
%\hline
Target & Figure & XPath\\
% & Figure & \\
\hline
\hline
\multirow{2}{*}{Name} & Before & /html/div[1]/div[1]/div[1]/span \\
%\hhline{~--}  
& After & /html/div[1]/div[1]/div[1]/\circled{{\color{black}div}}/span \\
\hline
\multirow{2}{*}{Description} & Before & /html/div[1]/div[1]/div[2]/div \\
%\hhline{~--} 
& After  & /html/div[1]/div[1]/div[2]/div \\
\hline
\multirow{2}{*}{Info} & Before & /html/div[1]/a \\
%\hhline{~--} 
& After & /html/div[1]/a \\

\end{tabular}
}
\end{center}

 \caption{A vertically shift occurs by adding a parent node to the $span$ element. The XPath that reaches $Name$ must have a $div$ element added to maintain compatibility with the altered HTML tree.}
\label{fig:shiftvert}
\end{figure}

  Our method is based on the principle that all shifts can be broken down into a combination of vertical and horizontal shifts. We can take advantage of this by considering all tree permutations. Given a tree $T$ with nodes $t \in T$, each node is a sub-tree and has a parent $t.parent$ and  a set of children $t.children$. A tree contains many paths $p$ with elements $p_i \in T$, when the path travels down the tree: $p_{i+1} \in p_{i}.children$.

\begin{definition} 

A \textbf{vertical shift} is a tree modification where a node is inserted on the path from the root to the target element.  Formally, for some path  
$p =\{p_1, \ldots, p_i, p_{i+1}, \ldots p_n\},$ a vertical shift occurs when some new node, lets call $s$, is inserted and  
$$p' = \{p_1, \ldots, p_i, s, p_{i+1}, \ldots, p_n\}$$ or, if a node is removed, 
$$p' = \{p_1, \ldots, p_i, p_{i+2}, \ldots, p_n\}.$$
  
\end{definition} 

A vertical shift causes an insertion or removal in an XPath, see the example in Figure \ref{fig:shiftvert}. 

\begin{definition}

A \textbf{horizontal shift} is a tree modification where a sibling element of a node is inserted. Formally, for some path  
$p = \{p_1, \ldots, p_{i-1}, p_i,p_{i+1}, \ldots,p_n\},$ a horizontal shift occurs when some new node, lets call $s$ where $s \neq p_i$, is inserted 
$$p' = \{p_1, \ldots,  p_{i-1}, s,p_{i+1}, \ldots, p_n\}.$$ $p_i$ may still exist in the tree but a different node now connects the two path segments $p_1, \ldots,  p_{i-1}$ and $p_{i+1}, \ldots, p_n$.

\end{definition}

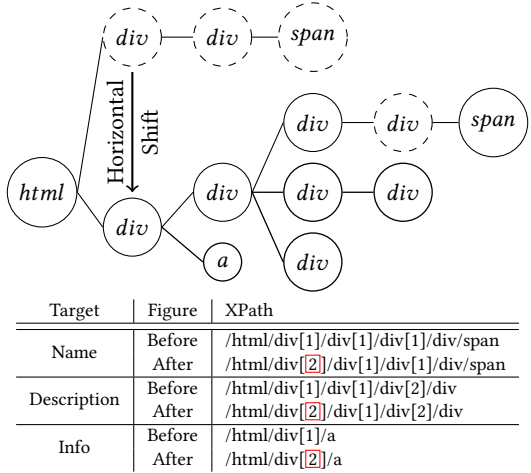
\begin{figure}[ht]

%\begin{center}

%{\scalefont{0.8}
	\begin{tikzpicture}[scale=0.8,
	leaf/.style={circle, draw=none, fill=black,
	        text centered, text=white},grow=right,
%	redfill/.style={fill=red!60},
	bold/.style={},
	notbold/.style={line width=0.5pt},
	sibling distance=33pt
	]
	%\draw (6.4,0) -- ++(0,-5);
	\begin{scope}
	\node[bold]  [circle,draw] (fa){$html$}
	%	\node [circle,draw] (f){$div_1$}
		child[bold]  {node [circle,draw,bold] (1a) {$div$}
			child[black, notbold]  {node [circle,draw] (aa) {$a$}}	
			child[bold]  {node [circle,draw,bold] (10a) {$div$}
	% 				<div id="4"></div>
					child[black, notbold]  {node [circle,draw] (4a) {$div$}}
	% 				<div id="9">
					child[black, notbold]  {node [circle,draw] (9a) {$div$}
	% 					<div id="2">div2</div>
						child {node [circle,draw] (2a) {$div$}}
					}
	% 				</div>
	% 				<div id="5">
					child[bold]  {node [circle,draw,bold] (5a) {$div$}
						child[bold]  {node [circle,draw,dashed,bold] (3a) {$div$}
		% 					<span id="6">span</span>
							child[bold]  {node [circle,draw,bold] (span1a) {$span$}}
						}
					}
	% 				</div>
				}
	% 			</div>
	% 			<a>a</a>
			}
			child {node [circle,draw,yshift=1.6cm,dashed] (11a) {$div$}
	% 				<div id="5">
					child {node [circle,draw,dashed] (12a) {$div$}
	% 					<span id="6">span</span>
						child {node [circle,draw,dashed] (span2a) {$span$}}
					}
			};

	\draw[thick,<-,style={sloped,anchor=north,auto=false}] (1.5,0) -- 
	node[above] {Horizontal} node[below] {Shift} (1.5,2);
			
	\end{scope}
	
	%\draw[<->,dashed,color=gray!100] (f) -- node[above,sloped] {} (fa);
	
	\end{tikzpicture}
%	}
%}
\begin{center} 
{\scalefont{0.8}
\begin{tabular}{c|c|l}
%\hline
Target & Figure & XPath\\
% & Figure & \\
\hline
\hline
\multirow{2}{*}{Name} & Before & /html/div[1]/div[1]/div[1]/div/span \\
%\hhline{~--}  
& After & /html/div[\circled{{\color{black}2}}]/div[1]/div[1]/div/span \\
\hline
%\hline
\multirow{2}{*}{Description} & Before & /html/div[1]/div[1]/div[2]/div \\
%\hhline{~--} 
& After  & /html/div[\circled{{\color{black}2}}]/div[1]/div[2]/div \\
\hline
%\hline
\multirow{2}{*}{Info} & Before & /html/div[1]/a \\
%\hhline{~--} 
& After & /html/div[\circled{{\color{black}2}}]/a \\
%\hline
\end{tabular}
}
\end{center}

\caption{A sibling $div$ element is added under the $html$ parent. This is a horizontal shift that affects many XPaths at once due to the change of a child index from $div[1]$ to $div[2]$.}
\label{fig:shifthori}
\end{figure}

A horizontal shift causes the index of a node to change, see the example in Figure \ref{fig:shifthori} .

With formal definition of shifts, we study a large dataset of 117,422 pages from 75 websites in 8 vertical markets %(Figure \ref{fig:compat-xpaths-per-domain})% 
described in Section \S \ref{sec:evaluation} to empirically analyze possible reasons for shifts. 

One indicator of shifts is when multiple XPaths are needed to extract the same attribute from different web pages of the same website.  The probability of a domain requiring multiple compatible XPaths decreases as the number of XPaths increases. Even with this good news, 116 out of the 231 attributes have more than one XPath associated with them. The most difficult domain and attribute are {\em barnesandnoble}'s {\em title} with 270 unique XPaths required. Inspecting the mean XPaths required for each domain we can observe slight chunks which would imply possible clusters and maybe some similarities between the websites. We identify three main possible reasons for shifts: 

%\begin{figure}[htp]
%\begin{center}
  %\includegraphics[width=1\columnwidth]{compat-xpaths-per-domain.png}
  
%  \vspace{-15pt}
  
%   \subfloat[\label{fig:compat:attributeall}]{%
%      \includegraphics[width=0.33\columnwidth]{xpath-count-attribute-all.png}
%    }%
%    \subfloat[\label{fig:compat:attributegt1}]{%
%      \includegraphics[width=0.33\columnwidth]{xpath-count-attribute-gt1.png}
%    }%
%    \subfloat[\label{fig:compat:domainall}]{%
%      \includegraphics[width=0.30\columnwidth]{xpath-count-domain-all.png}
%    }

 % \caption{
%Number and distribution of compatible XPaths are shown for attributes (\ref{fig:compat:attributeall} and \ref{fig:compat:attributegt1}) and domains (\ref{fig:compat:domainall}). \ref{fig:compat:attributeall} %includes the entire dataset including the ones with no shift problems to show the overall distribution. \ref{fig:compat:attributegt1} shows the distribution of attributes that have been shifted (because they have %more than one compatible XPath). The mean number of compatible XPaths for each domain is shown in \ref{fig:compat:domainall}.
%}

%	
%  \label{fig:compat-xpaths-per-domain}
%\end{center}
%\end{figure}

\begin{itemize}
  
  \item \textbf{Inconsistent templates} : A website might present items to a user differently depending on a property of that item. For example, an item on sale may have a graphic inserted which shifts the DOM. In our dataset, \url{collegeboard.com} uses different templates for public and private universities that results in shifts.
  
  \item \textbf{Temporal changes} : Over time the developer may change the site to fix a bug, add a feature, or perform a redesign. Changes can be related to user tracking, advertising, or updated template software.
  
  \item \textbf{DOM cleaner inconsistencies} : A DOM cleaner (details in Section \S \ref{sec:workingwithdata}) needs to make assumptions when converting semi-structured HTML into XHTML. If the HTML is ambiguous this process will result in a DOM tree that does not match the intended DOM and will appear as a shift.

\end{itemize}

In our empirical study, we observe large groups of unique XPaths are due to inconsistent templates and small groups of unique XPaths (less than 4) are usually due to DOM cleaner inconsistencies.

Our next step is to mathematically quantify the difficulty in maintaining a wrapper in order to measure the disorder of the domain. 

\begin{definition}

\textbf{Attribute difficulty} represents the disorder of the attribute locations with respect to XPath annotations in a sample set of pages from a domain. The probability that a particular XPath for an attribute will be compatible with pages from a domain given a sample of pages is formalized as: $p(x) = \frac{\text{compatible pages}}{\text{total pages}}$. We define attribute difficulty in Eq \ref{eq:attributeentropy}: 

\begin{equation}
H(\text{attribute}) = 
-\displaystyle\sum_{xpaths} p(xpath) \log  p(xpath)
\label{eq:attributeentropy}
\end{equation}
\end{definition}

When looking now at Table \ref{tab:uniquexpaths} the attribute difficulty can be used to quantify the difference between the two attributes presented for the domain \textbf{deepdiscount}. Intuitively {\em title} appears more difficult than {\em author} due to the long list of unique XPaths. Attribute difficulty confirms this with 0.93 for {\em title} and 0.65 for {\em author}. One strength of the difficulty analysis is that it weights each XPath to take into account outliers that are only compatible with 1 or 2 webpages. This is important because these outliers will not significantly impact accuracy and therefore should not impact the difficulty.

\begin{table*}
%\resizebox{\linewidth}{!}{%
\caption{The number of compatible pages in each domain for each XPath. For the XPath listed, the ``\# Compatible'' is the number of pages that have data stored at the location specified by that XPath.}
\label{tab:uniquexpaths}
{\small
\begin{tabular}{ c c l }
\toprule
Domain/ &  \#& \\
Attribute &   Compatible & Unique XPath\\
\hline
\hline
& 79 & /html/body/div/div[2]/div[1]/ul/li/ul/li/ul/li/ul/li/ul/li/ul/li/span \\
%\hhline{~--}
& 44 & /html/body/div/div[2]/div[1]/ul/li/ul/li/ul/li/ul/li/ul/li/ul/li/ul/li/span \\
%\hhline{~--}
deepdiscount/ & 20 & /html/body/div/div[2]/div[1]/ul/li/ul/li/ul/li/ul/li/ul/li/ul/li/ul/li/ul/li/span \\
%\hhline{~--}
title& 9 & /html/body/div/div[2]/div[1]/ul/li/ul/li/ul/li/ul/li/ul/li/ul/li/ul/li/ul/li/ul/li/span \\
%\hhline{~--}
& 577 & /html/body/div/div[2]/div[2]/div/div/div/div/div[1]/div/h1 \\
%\hhline{~--}
& 1237 & /html/body/div/div[2]/div/div/div/div[1]/div/div[1]/div/h1 \\

\hline
\hline
deepdiscount/& 738 & /html/body/div/div[2]/div[2]/div/div/div[1]/div/div[2]/div[1]/div[2]/div/ul/li[1]/a \\
%\hhline{~--}
author & 1257 & /html/body/div/div[2]/div/div/div/div[1]/div/div[2]/div[2]/div/div[1]/div/ul/li[1]/a[1] \\
\bottomrule
\end{tabular}
}
%}
\end{table*}

\begin{definition}

The \textbf{domain difficulty} is the mean difficulty of it's attributes. 

\end{definition}

\begin{figure}[htp]
\begin{center}
\vspace{-10pt}
 \includegraphics[width=1.0\columnwidth]{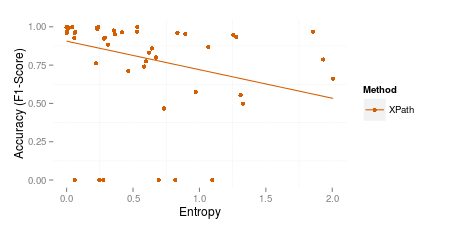}
 \caption{
	The domain difficulty versus the evaluated F1-Score of an XPath wrapper for every domain in our dataset. A linear trend-line is drawn to show the trend of the relationship.
	}
  \label{fig:entropyXpath}
  \vspace{-10pt}
\end{center}
\end{figure}

Inspecting domain difficulty can provide insight into where the current XPath method fails to solve the problem. In Figure \ref{fig:entropyXpath} we look at the F1-Score (measure for classification accuracy; the higher the better, explained in \S \ref{sec:evaluation}) versus the domain difficulty (a low value means less shifts occur). The plot reinforces the intuition that XPath usually works well on domains with low difficulty. Also we can confirm that when the difficulty is high XPath does not perform as well. This makes sense because higher difficulty means that there is more disorder in the set of XPaths for that domain which causes them to fail.

\begin{figure}[t]
\begin{center}
\vspace{15pt}
\begin{tikzpicture}
\node [text width=0.3\textwidth,yshift=+35pt]{
{
\scalefont{0.90}
\begin{Verbatim}[commandchars=\\\{\},codes={\catcode`$=3\catcode`^=7\catcode`_=8}]
<html>
<div>
\color{red}    <div>
\color{red}        <div>
\color{red}            <span>Sale!</span>
\color{red}        </div>
\color{red}    </div>
    <div>
        <div>
\color{red}            <div>
                <span>Name</span>
\color{red}            </div>
        </div>
        <div>
            <div>Desc</div>
        </div>
        <div>Stock</div>
    </div>
    <a>Info</a>
</div> 
</html>
\end{Verbatim}
}
};
\begin{scope}[text width=59pt,xshift=110pt,yshift=+90pt]

\node[yshift=-35pt]{\includegraphics[width=70pt]{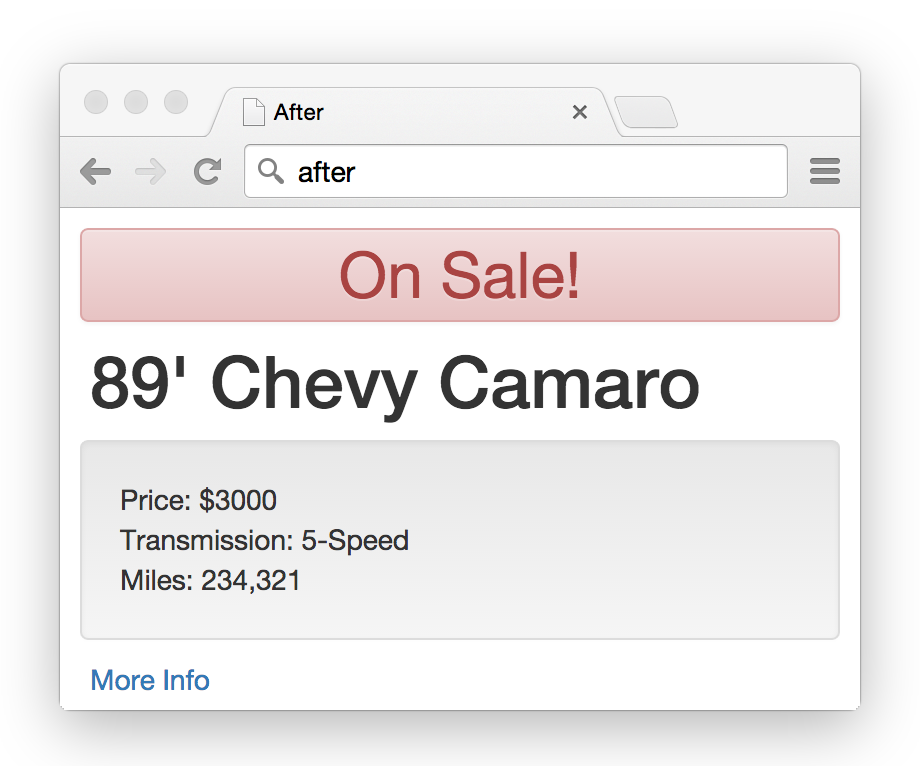}};

\end{scope}

\draw[thick,->,style={sloped,anchor=north,auto=false}] 
(0.5,2.3) --  node[above] {Rendered} (2.5,2.3);

\begin{scope}[scale=0.7,
leaf/.style={circle, draw=none, fill=black,
        text centered, text=white},grow=right,
redfill/.style={fill=red!60},
bold/.style={},
notbold/.style={line width=0.5pt},
yshift=-190pt,
xshift=-50pt,
	sibling distance=33pt	
]

\draw[thick,->,style={sloped,anchor=north,auto=false}] 
(-2,2.5) --  node[above] {DOM View} (0,1);	

\node[bold]  [circle,draw,bold] (fa){$html$}
%	\node [circle,draw] (f){$div_1$}
	child[bold]  {node [circle,draw,bold] (1a) {$div$}
		child[black, notbold]  {node [circle,draw] (aa) {$a$}}
		child[bold]  {node [circle,draw,bold] (10a) {$div$}
% 				<div id="4"></div>
				child[black, notbold]  {node [circle,draw] (4a) {$div$}}
% 				<div id="9">
				child[black, notbold]  {node [circle,draw] (9a) {$div$}
% 					<div id="2">div2</div>
					child {node [circle,draw] (2a) {$div$}}
				}
% 				</div>
% 				<div id="5">
				child[bold]  {node [circle,draw,bold] (5a) {$div$}
					child[bold]  {node [circle,draw,dashed,bold, redfill] (3a) {$div$}
	% 					<span id="6">span</span>
						child[bold]  {node [circle,draw,bold] (span1a) {$span$}}
					}
				}
% 				</div>
			}
% 			</div>
% 			<a>a</a>
		}
		child {node [circle,draw,yshift=1.6cm,dashed, redfill] (11a) {$div$}
% 				<div id="5">
				child {node [circle,draw,dashed, redfill] (12a) {$div$}
% 					<span id="6">span</span>
					child {node [circle,draw,dashed, redfill] (span2a) {$span$}}
				}
		};
\end{scope}

%\draw[<->,dashed,color=gray!100] (f) -- node[above,sloped] {} (fa);

\end{tikzpicture}

{\scalefont{0.8}
\begin{tabular}{c|l}
%\hline
Target &XPath\\
% & Figure & \\
\hline
\hline
\multirow{1}{*}{Name} &/html/div[2]/div[1]/div[1]/div/span \\
\hline
\multirow{1}{*}{Description} & /html/div[2]/div[1]/div[2]/div \\
\hline
\multirow{1}{*}{Info} &  /html/div[2]/a \\
%\hline
\end{tabular}
}
\end{center}
	
%}
\caption{The starting webpage has simulated shift applied by the addition of the words ``On Sale!''. The nodes used to add this text are designed to break existing methods. These are identified with dashed borders. The path of nodes from the root to $Name$ node are highlighted in red.}
\label{fig:examplethreeafter}
\end{figure}

%for s in `ls`; do echo $s; for i in `ls $s`; do echo $i; cat $s/$i/**/1.item |grep : | sort | uniq | awk -F':' '{print $1}' | uniq -c; done ; done

\section{XTreePath (XPath+TreePaths)}

We now utilize the knowledge gained from our shift analysis that more attribute difficulty means lower performance of XPath annotations. 
%Entropy is a symptom of shift so if our method can properly take shift into account then it will increase accuracy.  
Then we will explain how our Recursive Tree Matching algorithm searches node by node using similarity of sub-trees to train examples to simultaneously accommodate for vertical and horizontal shifts. In order to store the relevant data for searching we first explain tree paths.

\subsection{TreePaths}
\label{sec:tpath}
We seek for an efficient wrapper method that can fix incompatible wrappers automatically when shifts happen. In order to have enough information to fix wrappers in an automated way we extract not only the direct indexing into the document but also the context of those elements.  Instead of consulting the entire training set to repair a wrapper we store the tree structure immediately surrounding the target data. The algorithm starts building the tree path at the least common ancestor (LCA) of the target elements.

\begin{definition}

\textbf{Least common ancestor (LCA) of target elements:} The least common ancestor is an element that exists on every path from the root to each target element. The LCA is unique in that there is no other common element that is closer to every target element.

\end{definition}

\begin{definition}

A \textbf{tree path} is identified by $\tau$. It is a sequence of trees in an HTML Document Object Model (DOM) starting from the least common ancestor ($\tau_0$) and ending at the target element ($\tau_n$) as follows: $$TreePath = \tau = \langle\tau_0, \tau_1, \ldots, \tau_n\rangle, \tau_i \in DOM$$

\end{definition}

A tree path is an extension of the XPath concept. With XPath, the elements of the path are tag names that describe the sequence to the target. In contrast, a tree path includes an entire sub-tree starting from each element in the sequence to the target. This is to provide sufficient contextual information from which each element was extracted to aid in the wrapper recovery later. An example tree path of length four is shown in Figure \ref{fig:TPath}.

Extracting tree paths from an HTML DOM is shown in Algorithm \ref{alg:buildTPath}. First, we find the least common ancestor (LCA) between all the labeled elements. Next, starting from the target element, each element is added to a vector and then it's parent element and so on until the LCA is reached.  Next, we add the LCA because it was not added in the above loop. Finally the elements are reversed and returned so that they start with the LCA and end with the target element.

\begin{algorithm}
\LinesNumbered
\caption{Build Tree Path From Training DOM}
\label{alg:buildTPath}
\KwIn{Training DOM $dom$ \\
\hspace{27pt}Labeled elements $E=\{e_1,\ldots,e_k\}$ \\
\hspace{27pt}Target element $e$}
\KwOut{Tree Path $\tau$}

$dom_{LCA} = LCA(E, dom)$
\label{alg:buildTPath:prefix}

\While{ $dom_{LCA} \neq e$}{

$\tau.add(e)$
\label{alg:buildTPath:add}
$e = e.getParent()$

}
$\tau.add(e.getParent())$
\label{alg:buildTPath:addprefix}

return $ \tau^R$
\label{alg:buildTPath:reverse}
\end{algorithm}

\subsection{Recursive Tree Matching}
\label{sec:rtm}

We learned from analyzing shifts causing incompatibility that the majority are composed only a very small number of vertical and horizontal shifts. With our proposed Recursive Tree Matching, we jump over these shifts by matching sub-trees on each side of the shift. The LCA, which is the root of the tree path, provides a starting point and allows us to ignore shifts that have occurred outside where the target data is. Starting here also allows us to reduce the complexity of the search. The proceeding elements of the tree path are matched to their most similar nodes in order to align trees that existed previously unshifted. The objective function is shown in Eq \ref{eq:rtm}. Here the maximum matching sequence of $e$ (DOM elements) to the data contained in each $\tau_i$ is found. This maximization is iterative with constraints which requires two maximization sections.

\begin{equation}
max \left(\sum_{\tau_i \in \tau}
\{e_{i+1} = argmax_{e\in e_{i}}(match(\tau_i, e))\}\right)
\label{eq:rtm}
\end{equation}

The algorithmic detail, including the dynamic programming heuristic function, of Recursive Tree Matching is shown in Algorithm \ref{alg:xpathandrtm}. In this pseudo code a reference to an element of the DOM is kept as $d$ and updated as the search progresses. Line \ref{alg:useTPath:max} is the core where each element of the tree path $\tau_i$ is matched to its most similar DOM element in $d$.  HTML Tree similarity is calculated using a modified Simple Tree Matching algorithm \cite{yang_identifying_1991} which is designed to deal with HTML specifically by taking into account the class, style, id, name attributes of each node. If a max similarity is 0 then the element is considered not found. Using this method we perform a heuristic search through the tree using concise information from the training data.

\begin{algorithm}[!ht]
\LinesNumbered
\caption{Wrapper Induction (Recursive Tree Matching)}
\label{alg:xpathandrtm}
\KwIn{Tree path $\tau$ \\ \hspace{30pt}HTML DOM $dom$  }
\KwOut{Resulting $data$}

$d \leftarrow dom$

\For{$\tau_i \in \tau $ }{

    \For{$e \in d$}{
    
    $d = argmax_e(html\_tree\_match(\tau_i,e))$
    
    \label{alg:useTPath:max}
    
    \commentcode{1}{//If match is 0 then not found}

    }

}

return $d$

\vspace{10pt}

$html\_tree\_match(a,b):$

\If{a and $\tau_i$ contain distinct symbols}{

return $0$

}\Else{

$m \leftarrow $ the number of first-level sub-trees of $a$

$n \leftarrow $ the number of first-level sub-trees of $b$

$M[i,0] \leftarrow 0 $ for $ i = 0,\ldots,m$

$M[0,j] \leftarrow 0 $ for $ j = 0,\ldots,n$

\For{$i = 1$ to $m$}{

	\For {$i = 1$ to $n$}{

		$x \leftarrow M[i,j-1]$

		$y \leftarrow M[i-1,j]$

		$z \leftarrow M[i-1,j-1]+ html\_tree\_match(a_i,b_j)$

		$M[i,j]	\leftarrow max(x,y,z)$
	}
}

\For{$attr \in \{class,style,id,name\}$}{
	\If{$a_{attr} == b_{attr}$}{
		$attrMatch \leftarrow attrMatch + 0.25$;
	}
}

return $M[m,n] + (attrMatch * 0.5) + 1$
}

\end{algorithm}

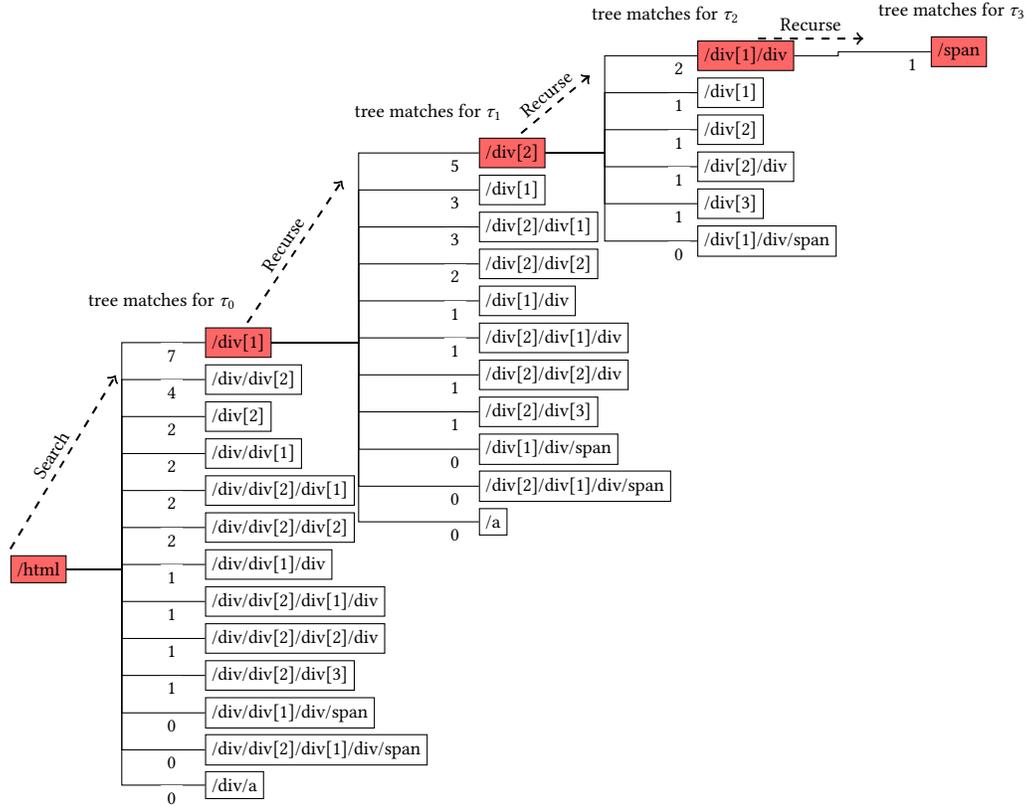
\begin{figure*}[t]
\begin{center}

{\scalefont{0.8}
\begin{tikzpicture}[scale=.7, sibling distance=20pt, edge from parent fork
right]
%, edge from parent fork right
\tikzset{grow'=right}
\tikzset{level 1/.style={level distance=90pt}}
\tikzset{level 2/.style={level distance=130pt}}
\tikzset{level 3/.style={level distance=100pt}}
\tikzset{level 4/.style={level distance=100pt}}
%\tikzset{execute at begin node=\strut}
\tikzset{every node/.style={anchor=base west,fill=white!100}}

\node[draw,fill=red!60] (fa) {/html} 
	child {node[draw,fill=red!60] (1a) {/div[1]}
		child {node[draw,fill=red!60] (10a) {/div[2]}
			child {node[draw,fill=red!60] (5a) {/div[1]/div}
				child {node[draw,fill=red!60] (span1a) {/span}
					edge from parent node[auto=right,pos=.9] {1} 
					node[auto=left,pos=.9, yshift=10pt,xshift=15pt] {tree matches for 
					$\tau_3$}} edge from parent node[auto=right,pos=.9] {2}
				node[auto=left,pos=.9, yshift=10pt, xshift=-5pt] {tree matches for 
				$\tau_2$};} child {node[draw]{/div[1]}
				edge from parent node[auto=right,pos=.9] {1};}
			child {node[draw]{/div[2]}
				edge from parent node[auto=right,pos=.9] {1};}
			child {node[draw]{/div[2]/div}
				edge from parent node[auto=right,pos=.9] {1};}
			child {node[draw]{/div[3]}
				edge from parent node[auto=right,pos=.9] {1};}
			child {node[draw]{/div[1]/div/span}
				edge from parent node[auto=right,pos=.9] {0};}
			edge from parent node[auto=right,pos=.9] {5}
			node[auto=left,pos=.9, yshift=10pt,xshift=-10pt] {tree matches for $\tau_1$};}
		child {node[draw]{/div[1]} 
			edge from parent node[auto=right,pos=.9] {3};}
		child {node[draw]{/div[2]/div[1]} 
			edge from parent node[auto=right,pos=.9] {3};}
		child {node[draw]{/div[2]/div[2]} 
			edge from parent node[auto=right,pos=.9] {2};}
		child {node[draw]{/div[1]/div} 
			edge from parent node[auto=right,pos=.9] {1};}
		child {node[draw]{/div[2]/div[1]/div} 
			edge from parent node[auto=right,pos=.9] {1};}
		child {node[draw]{/div[2]/div[2]/div} 
			edge from parent node[auto=right,pos=.9] {1};}
		child {node[draw]{/div[2]/div[3]} 
			edge from parent node[auto=right,pos=.9] {1};}
		child {node[draw]{/div[1]/div/span} 
			edge from parent node[auto=right,pos=.9] {0};}
		child {node[draw]{/div[2]/div[1]/div/span} 
			edge from parent node[auto=right,pos=.9]{0};} 
		child {node[draw]{/a} 
			edge from parent node[auto=right,pos=.9] {0};}
	edge from parent node[auto=right,pos=.8] {7}
		node[auto=left,pos=.9, yshift=10pt,xshift=-10pt] {tree matches for 
		$\tau_0$};} child {node[draw]{/div/div[2]}
	edge from parent node[auto=right,pos=.8] {4};}
	child {node[draw]{/div[2]}
	edge from parent node[auto=right,pos=.8] {2};}
	child {node[draw]{/div/div[1]}
	edge from parent node[auto=right,pos=.8] {2};}
	child {node[draw]{/div/div[2]/div[1]}
	edge from parent node[auto=right,pos=.8] {2};}
	child {node[draw]{/div/div[2]/div[2]}
	edge from parent node[auto=right,pos=.8] {2};}
	child {node[draw]{/div/div[1]/div}
	edge from parent node[auto=right,pos=.8] {1};}
	child {node[draw]{/div/div[2]/div[1]/div}
	edge from parent node[auto=right,pos=.8] {1};}
	child {node[draw]{/div/div[2]/div[2]/div}
	edge from parent node[auto=right,pos=.8] {1};}
	child {node[draw]{/div/div[2]/div[3]}
	edge from parent node[auto=right,pos=.8] {1};}
	child {node[draw]{/div/div[1]/div/span}
	edge from parent node[auto=right,pos=.8] {0};}
	child {node[draw]{/div/div[2]/div[1]/div/span}
	edge from parent node[auto=right,pos=.8] {0};}
	child {node[draw]{/div/a}
	edge from parent node[auto=right,pos=.8] {0};};

 	\draw[thick,->,style={dashed,sloped,anchor=north,auto=false}] 
 	(0,0.5) --  node[above] {Search} (2,3.8);
 	
	\draw[thick,->,style={dashed,sloped,anchor=north,auto=false}] 
 	(4.5,4.8) --  node[above] {Recurse} (6.3,7.5);
 	
 	\draw[thick,->,style={dashed,sloped,anchor=north,auto=false}] 
 	(9.7,8.4) --  node[above] {Recurse} (11,9.5);
 	
 	\draw[thick,->,style={dashed,sloped,anchor=north,auto=false}] 
 	(14.2,10.2) --  node[above] {Recurse} (16.2,10.2);
	
\end{tikzpicture}
}

\caption{The search tree when performing Recursive Tree Matching on Figure
\ref{fig:examplethreeafter} using the tree path in Figure \ref{fig:TPath}.
Each transition represents a search using the next tree of the tree path.
Each box displays represents a sub-tree offset from the source of the transition. Inside the box displays the similarity result of Simple Tree Matching and the XPath leading to the sub-tree in the form similarity:XPath.
The result of this recursive tree matching search yields $/html/div/div[2]/div[1]/div/span$ which is quite different from our starting XPath of $/html/div/div[1]/div[1]/span$.}
\label{fig:searchtree}
\end{center}
\vspace{-10pt}
\end{figure*}

A demonstration of the recursive tree matching process is shown in Figure \ref{fig:searchtree}. The shifted HTML DOM presented in Figure \ref{fig:examplethreeafter} is searched using the Recursive Tree Matching method with the tree path we extracted from original DOM in Figure \ref{fig:examplethreeafter} shown in Figure \ref{fig:TPath}. We first start by trying to directly look up the target data using the original sequence of elements. This results in a failure causing the algorithm to perform wrapper recovery.

Recovery starts by searching every element in the HTML DOM for the sub-tree that has the highest similarity to $\tau_0$ (the LCA). In Figure \ref{fig:searchtree} the element $/div$ has a similarity score of 7 which is higher than all other sub-trees. The similarity is calculated using the $html$\_$tree$\_$match$ method. The score of 7 is calculated as the maximum overlap of one tree with the other given the liberty of horizontal shifts. 
%In Figure \ref{fig:overview} the intersection where $\tau_0$ can overlap $/div[1]$ by at most 7 nodes. In this search our method has avoided the incorrect $/div[2]$ and navigated the tree using the context of the choices.

Once we have focused on $/div[1]$, this sub-tree is now searched using the second element of the tree path. A similarity score of 5 yields $/div[2]$ as the most similar sub-tree. Next, the algorithm will find a most similar sub-tree by jumping over the element $/div[1]$ to find $/div[1]/div$ has a higher similarity. This search will result in the $/span$ element being located and the wrapper successfully repaired. This example showcases the power of Recursive Tree Matching method in dealing with addition of identical trees and the extension of trees.

\subsection{Complexity}
\label{sec:complexity}

The worst case complexity of the recursive tree matching method is $O(|\tau|n_1n_2)$ where $n_1$ and $n_2$ are the number of elements in the training and test DOM trees. This is derived from the Simple Tree Matching (STM) complexity, which is reduced using dynamic programming, being $O(n_1n_2)$. There are $|\tau|$ iterations of the algorithm, each needing to perform an STM search. The cost at each iteration will be smaller than the previous but for this analysis we round up. Also, in our method we reduce the initial size of $n_1$ by selecting the LCA instead of the root element. Empirically the initial $n_1$ value is very small, about 30.

\section{Evaluation}
\label{sec:evaluation}

In this section we aim to show that XTreePaths can be utilized to complement and outperform the existing dominant method XPath. The method presented has no parameters that require tuning. Our goal is to design robust and practical method which can be easily extended under different scenarios. 

In order to test the robustness of the methods, the percentage of the dataset used for training is varied. This allows a comprehensive comparison between the following three methods:

\begin{itemize}
  \item \textbf{XPath} : Each training example has an XPath to be the target node that is combined into a set of possible paths. Each path is attempted on the testing examples until there is a valid path.
  \item \textbf{XTreePath} : Firstly XPaths are attempted. If it is not successful then a tree paths is used to attempt recovery.
  \item \textbf{TreePath} : Only a tree path is used without XPath. A tree path is extracted from each training example starting from the LCA of the target elements for that domain. Then Recursive Tree Matching is used to search the DOM tree.
  \item \textbf{ScrapingHub} : The web service offered at \url{scrapinghub.com} is used as a blackbox to evaluate the state of the art offered commercially by industry. (No authors are affiliated with ScrapingHub)
\end{itemize}

To compare XTreePath and XPath we use a large dataset built by Qiang Hao \cite{hao_one_2011} to benchmark per-vertical wrapper repair instead of per-domain. This dataset contains a total of 117,422 pages from 75 diverse websites in 8 vertical markets that covers a broad range of topics from university rankings to NBA players. The composition is displayed in Table \ref{table:datasets}. For each vertical market a set of (3-5) common attributes are labelled on every page.  We make our data and code available for comparison at \url{http://kdl.cs.umb.edu/w/datasets/}.

\begin{table*}[ht]
\caption{Composition of the dataset. Domains in each vertical are shown with the number of sample instances. }
\label{table:datasets}
\resizebox{\linewidth}{!}{%
\begin{tabular}{c c c p{18.0cm}}
\toprule
Vertical & \#Sites & \#Pages & Domains\\
\midrule
Autos & 10 & 17,883 & 
aol,autobytel,automotive,autoweb,carquotes,cars,kbb,motortrend,msn,yahoo \\
%\hhline{~---}
Books & 10 & 15,990 & 
abebooks,barnesandnoble,bookdepository,booksamillion,borders,christianbook,deepdiscount,waterstones\\
%\hhline{~---}
Cameras & 10 & 6,991  & 
amazon,beachaudio,buy,compsource,ecost,jr,newegg,onsale,pcnation,thenerds \\
%\hhline{~---}
Jobs & 10 & 19,963 &
careerbuilder,dice,hotjobs,job,jobcircle,jobtarget,monster,nettemps,rightitjobs,techcentric\\
%\hhline{~---}
Movies & 10 & 16,000  &
allmovie,amctv,boxofficemojo,hollywood,iheartmovies,imdb,metacritic,rottentomatoes\\
%\hhline{~---}
NBA Players & 9 & 3,966 &
espn,fanhouse,foxsports,msnca,nba,si,slam,usatoday,wiki \\
%\hhline{~---}
Restaurants & 10 & 19,928 &
fodors,frommers,gayot,opentable,pickarestaurant,restaurantica,tripadvisor,urbanspoon,usdiners,zagat \\
%\hhline{~---}
Universities & 10 & 16,701 &
collegeboard,collegenavigator,collegeprowler,collegetoolkit,ecampustours,embark,matchcollege,princetonreview,studentaid,usnews\\
%\hline%
\bottomrule
\end{tabular}
}

\end{table*}

\subsection{Working With Data}
\label{sec:workingwithdata}

It is important to document the difficulties of dealing with datasets in this field in order to ensure that the benchmark datasets used here can be utilized by other researchers. 

The main issue is that HTML in the wild does not always map to the same DOM representation, it is highly dependent on the HTML Cleaner used.  There is no standard mapping to convert HTML into a properly formatted XHTML file.  There are common algorithms used by browsers but there is no agreed specification of how they convert HTML to XHTML. Different cleaners convert HTML in different ways leading to an incompatibility of XPath annotations. A training set made with one cleaning engine will not work using another engine. 

The libraries used for our work were chosen as the most reliable and competent methods after a comparative study on major libraries was performed. The complete list of libraries we evaluated are labeled in Table \ref{tab:libs} as having the following properties:

\begin{itemize}
  \item \textbf{Cleaning} - These are used to convert from a non-standard HTML file to an XHTML file.  The corrections include tag closing, name-space filtering, and tag nesting.  This is required because most HTML on the Internet is non-standard.
  \item \textbf{Representation} - This library provides things such as DOM
  traversal, insertion, and removal.  This library is used to build sub-trees and represent namespaces. Most of these libraries are not tolerant to non-standard HTML and require cleaning before they can turn HTML into a DOM.
  \item \textbf{Query} - These methods can include XPath, XQuery, XML-GL, or XML-QL.
\end{itemize}

In this paper, our research is done using the well supported open source libraries JTidy and Dom4j. These libraries are written in Java and support multi-threading. JTidy's performed consistently for cleaning and intergraded cleanly into Dom4j. Dom4j has a clean query interface to lookup using XPath as well as a clean Representation interface for implementing tree paths and recursive tree matching. Other Java libraries such as JSoup and TagSoup are designed for their own query language instead of XPath and exposing a DOM.

A few pages today retrieve their content using JavaScript once the page is loaded.  This means the HTML retrieved with the initial GET request does not contain the full product information.  A way to solve this problem is to use a library that will run the JavaScript on the page or to scrape the data using a browser after it has run the JavaScript code.  It is better to get JavaScript out of the way during scraping due to the need to make AJAX calls to retrieve data that may be missing at a later date. For this reason we retrieve the pages using the FireFox web browser which will evaluate JavaScript as expected by the web developer.

%\definecolor{Gray}{gray}{0.85}
\definecolor{LightCyan}{rgb}{0.78,1,1}

\begin{table}[!h]
\label{table:classifications}
\begin{center}
\caption{Information Extraction Library Classification}
\label{tab:libs}
\begin{tabular}{c c c c c}
\toprule
Name & Lang & Clean & Rep. & Query\\
\midrule
JSoup & Java & $\star$ & $\star$ & $\star$\\
%\hline
NokoGiri & Ruby & $\star$ & $\star$ & $\star$\\
%\hline
TagSoup & Java & $\star$ & $\star$ & $\star$\\
%\hline
Taggle & C++ & $\star$ & $\star$ & $\star$\\
%\hline
Rubyful Soup%\footnote{\url{http://www.crummy.com/software/RubyfulSoup/}} 
& Ruby
&$\star$ & $\star$ & $\star$\\
%\hline
Beautiful Soup %\footnote{\url{http://www.crummy.com/software/BeautifulSoup/}}
&Python&$\star$ & $\star$ & $\star$\\
%\hline
NekoHTML %\footnote{http://nekohtml.sourceforge.net/}  
& Java & $\star$ & $\star$&$\star$\\
%\hline
Xom  %\footnote{http://www.xom.nu/} 
& Java & & $\star$ & $\star$\\
%\hline
Saxon %\footnote{http://saxon.sourceforge.net/} 
& Java & & $\star$ &$\star$\\
%\hline
Xerces %\footnote{http://xerces.apache.org/}
& Java & & $\star$ &$\star$\\
%\hline
HTMLParser %\footnote{http://htmlparser.sourceforge.net/} 
& Java & & $\star$&$\star$\\
%\hline
XStream %\footnote{http://xstream.codehaus.org/} 
& Java & & $\star$&$\star$\\
%\hline
\rowcolor{LightCyan} Dom4j %\footnote{http://dom4j.sourceforge.net/} 
& Java & & $\star$ & $\star$\\
%\hline
HTML Tidy %\footnote{http://tidy.sourceforge.net/} 
& C & $\star$ & &\\
%\hline
\rowcolor{LightCyan} JTidy %\footnote{http://jtidy.sourceforge.net/} 
& Java & $\star$ & &\\
%\hline
Tika %\footnote{http://tika.apache.org/} 
& Java & $\star$ & &\\
%\hline
HTMLCleaner %\footnote{http://htmlcleaner.sourceforge.net/} 
& Java & $\star$ & &\\
%\hline
Jaxen %\footnote{http://jaxen.codehaus.org/} 
& Java & & &$\star$\\
%\hline
Xalan %\footnote{http://xml.apache.org/xalan-j/} 
& Java & & &$\star$\\
\bottomrule

\end{tabular}
\end{center}
\end{table}

We use the standard machine learning metrics precision, recall, and F1-score as the evaluation metrics. In this domain, a true positive (tp) is an extraction of the correct data (verified using labeled data), a false positive (fp) is an extraction that resulted in the wrong data (something other than the correct data), and a false negative (fn) is an extraction that resulted in an error or ``not found''. Errors are caused by the lookup reaching a dead end. The following formulae are used: $precision = \frac{tp}{tp+fp}$, $recall = \frac{tp}{tp+fn}$, and $F1 = 2 \cdot \frac{precision \cdot recall}{precision + recall}$

\subsection{Difficulty Correlation}

\begin{figure}[ht]
\begin{center}
\vspace{-10pt}
  \includegraphics[width=1.0\columnwidth]{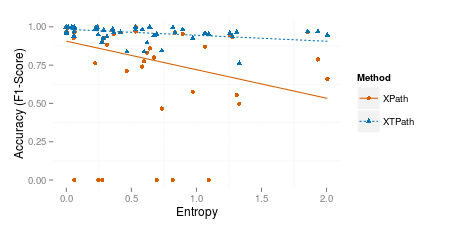}
  \caption{
The entropy of the domains is plotted against the F1-score of XTreePath and XPath. Linear trendlines are shown.
}
  \label{fig:entropyxtpathXpath}
  \vspace{-10pt}
\end{center}

\end{figure}

Using our new domain entropy measurement introduced in Section \S \ref{sec:shiftanalysis} we plot the entropy per domain over all vertical markets in Figure \ref{fig:entropyxtpathXpath} for XPath and XTreePath.  As the entropy increases XTreePath is able to maintain performance while XPath degrades. The higher a domain entropy value is, the more changes in HTML elements occur. XTreePath is more robust than XPath when dealing with shifts.

\subsection{Limiting Training Examples}

We evaluate the ability to recover from shift on each domain by splitting the pages of each domain into training and testing sets at various percentages. The pages of each domain are chosen randomly to simulate the different possible situations that could be encountered.

In the following experiments each method is trained on a percentage of the dataset. In these experiments; 10\% percentage trained means that only 10\% of the pages from a domain are used in training to build XPaths and XTreePaths. These are used to extract data from the remaining 90\% of the pages.

\begin{figure}[h]
\begin{center}
\vspace{-10pt}
\includegraphics[width=1.0\columnwidth]{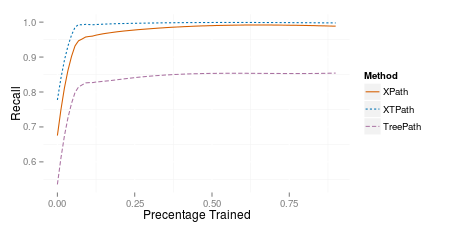}

\caption{Recall is not high when evaluating the TreePath alone. However, for XTreePath, when tree paths are used after XPath fails we are able to obtain a higher recall overall.
}
\vspace{-10pt}
\label{fig:varytraining:recall}
\end{center}
\end{figure}

In Figure \ref{fig:varytraining:recall} the aggregate recall is plotted against the percentage trained. This analyses how many wrappers are saved from needing to be relearned by using the different methods. We can observe the combination of XPath and tree paths as XTreePath achieves a significant increase which confirms they are complementing each other. This is important because these methods do not directly replace each other and together are able to provide a more robust data extraction pipeline. Also, it is important to note the largest increase is with a lower percentage of training data. This is desired because the algorithm can perform even if a small number of pages have been collected which is often the case. This is because every annotated training page is a cost to the system. Also, some data sources will increase in size over time causing the trained percentage will shrink over time.

\begin{figure}[h]
\begin{center}

\includegraphics[width=1.0\columnwidth]{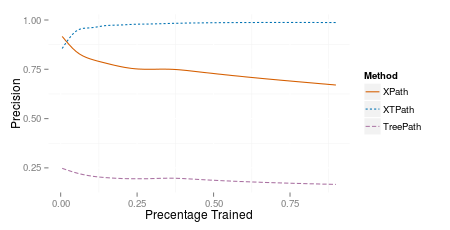}

\caption{Tree paths, by themselves, perform poorly but because there is only a small intersection between successful XPath and tree path extractions the true positives outweigh the false positives and drive the precision up.
}

\label{fig:varytraining:precision}
\end{center}
\end{figure}

Next we evaluate the aggregate precision in Figure \ref{fig:varytraining:precision}.  The most interesting result here is that as the training percentage is increased, the precision is reduced for XPath and tree paths.  As more examples are learned, XPath and tree path have more data to try which results in higher false positives. When the methods are combined in XTreePath the same number of false positives exists but the number of true positives increases and allows the precision formula to grow.

\begin{figure}[h]
\begin{center}

\vspace{-10pt}
\includegraphics[width=1.0\columnwidth]{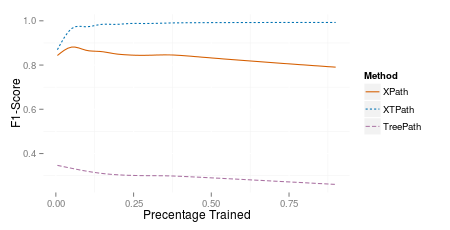}

\caption{XPaths and tree paths have slightly decreasing results due to precision. When combined they complement each other and increase as the training set size is increased because they perform better for recall.}
\vspace{-10pt}
\label{fig:varytraining:f1}
\vspace{-10pt}
\end{center}
\end{figure}

The aggregate F1-Score is shown in Figure \ref{fig:varytraining:f1}. Here the F1-Score of XTreePath consistently outperform XPath and tree paths alone. The advantage of XPath in precision is countered by it's low recall.

\subsection{Performance Per Vertical}

\begin{figure}[h]
\begin{center}
  \includegraphics[width=0.9\columnwidth]{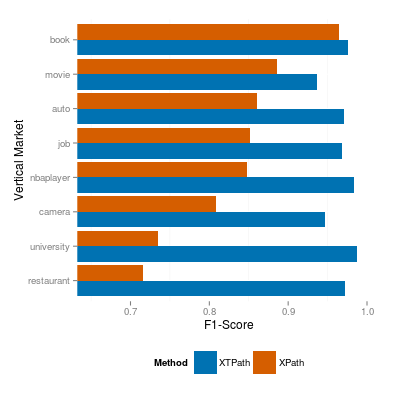}
  \caption[labelInTOC]{
Here we compare our method across eight vertical markets. XPath is sorted from
worst to best starting from the left.
}
  \label{fig:methodDatasetF1h}
\end{center}
\end{figure}

We are interested to know how the proposed XTreePath perform in vastly different web domains. In Figure \ref{fig:methodDatasetF1h} we analyze the mean F1-score of XPath and XTreePath in each vertical market. XTreePath performs consistently well against XPath in all vertical markets. We can draw from this analysis that book websites have more stable structure which allows XPaths to work consistently. We can also draw that restaurant and university sites have more dynamic structures with slight changes that can easily be accommodated for by XTreePath.

\subsection{Industry Baseline}

Finally, we compare our method to a current commercial solution, ScrapingHub, that tackles the same problem. This method is treated as a black box and we do not know how it works. In this evaluation both methods are trained on the same single example. Each method is then evaluated on the remaining examples from the domain. 

\begin{figure}[h]
\begin{center}
  \includegraphics[width=1\columnwidth]{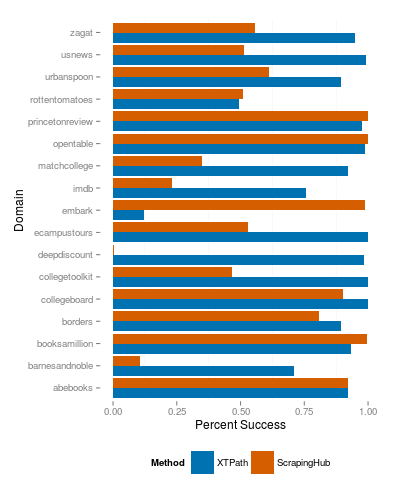}
  \caption[labelInTOC]{
Comparing XTreePath with ScrapingHub. One sample from each domain is used as train both systems. The methods are then evaluated for accuracy and shown here.
}
  \label{fig:scrapinghub}
\end{center}
\end{figure}

Figure \ref{fig:scrapinghub} shows the comparison using 17 randomly selected domains. XTreePath ties or beats ScrapingHub on 12/17 domains. For the domain {\em embark} the problems arise from two faults happening at the same time. First, an XPath fails when locating the address (a span[6] is shifted to a span[8]). XTreePath recovers this broken XPath but then fails on the phone number attribute (for which the learned XPath would have worked but the system was already trying to recover the wrapper). The weakness is that when all the children look the same the tree similarity doesn't work. This happened to be a perfect storm for XTreePath but would easily be fixed by adding another training example.

The {\em abebooks} results are identical. Why can't we improve this result? The weakness is that tree paths cannot deal with the shift in this site because it is confused with matching tree structure. The shifted site only has it's data rearranged but the tree structure has not changed. Even with using the node attributes it cannot repair the wrappers because they are also the same.

\section{Conclusion}

We have presented the XTreePath algorithm which is composed of XPath and tree paths together. Tree paths contain contextual information from training examples and are used by the recursive tree matching search algorithm. We have shown that with a simple data structure, the tree path, we can conquer shifts in webpages and reduce manual retraining. We evaluate our method on a massive and publicly available dataset where XTreePath consistently outperforms XPath alone.

A key advantage of the XTreePath method is that it complements existing methods and it does not need to replace them. We hope that this allows greater adoption in research and industry. Further work may utilize a semantic difference between trees by utilizing a cost matrix to weight differences in HTML elements unequally which would potentially increase accuracy but introduce additional parameters to test. To accelerate adoption we provide our easy to use implementation as open source and all the code necessary to evaluate it.

\section{Acknowledgements}
Partially funded by ThinkersR.Us, The University of Massachusetts Boston, The National Science Foundation Award Research Experiences for Undergraduates (NSF\#0755376), and The National Science Foundation Graduate Research Fellowship Program (Grant No. DGE-1356104). This work utilized the supercomputing facilities managed by the Research Computing Department at the University of Massachusetts Boston as well as the resources provided by the Open Science Grid, which is supported by the National Science Foundation and the U.S. Department of Energy's Office of Science.

\newpage
\bibliographystyle{ACM-Reference-Format}
\bibliography{infoextract} 


\begin{thebibliography}{20}

%%% the Web address.

\ifx \showCODEN    \undefined \def \showCODEN     #1{\unskip}     \fi
\ifx \showDOI      \undefined \def \showDOI       #1{#1}\fi
\ifx \showISBNx    \undefined \def \showISBNx     #1{\unskip}     \fi
\ifx \showISBNxiii \undefined \def \showISBNxiii  #1{\unskip}     \fi
\ifx \showISSN     \undefined \def \showISSN      #1{\unskip}     \fi
\ifx \showLCCN     \undefined \def \showLCCN      #1{\unskip}     \fi
\ifx \shownote     \undefined \def \shownote      #1{#1}          \fi
\ifx \showarticletitle \undefined \def \showarticletitle #1{#1}   \fi
\ifx \showURL      \undefined \def \showURL       {\relax}        \fi
% invisible to TeX
\providecommand\bibfield[2]{#2}
\providecommand\bibinfo[2]{#2}
\providecommand\natexlab[1]{#1}
\providecommand\showeprint[2][]{arXiv:#2}

\bibitem[\protect\citeauthoryear{Anton}{Anton}{2005}]%
        {Anton2005}
\bibfield{author}{\bibinfo{person}{Tobias Anton}.}
  \bibinfo{year}{2005}\natexlab{}.
\newblock \showarticletitle{{XPath-Wrapper Induction by generalizing tree
  traversal patterns}}.
\newblock \bibinfo{journal}{{\em Lernen, Wissensentdeckung und Adaptivitt
  (LWA)\/}} (\bibinfo{year}{2005}).
\newblock


\bibitem[\protect\citeauthoryear{Chidlovskii}{Chidlovskii}{2003}]%
        {Chidlovskii2003}
\bibfield{author}{\bibinfo{person}{Boris Chidlovskii}.}
  \bibinfo{year}{2003}\natexlab{}.
\newblock \showarticletitle{{Information extraction from tree documents by
  learning subtree delimiters}}.
\newblock \bibinfo{journal}{{\em IJCAI Workshop on Information Integration on
  the Web\/}} (\bibinfo{year}{2003}).
\newblock


\bibitem[\protect\citeauthoryear{Dalvi, Bohannon, and Sha}{Dalvi
  et~al\mbox{.}}{2009}]%
        {dalvi_robust_2009}
\bibfield{author}{\bibinfo{person}{Nilesh Dalvi}, \bibinfo{person}{Philip
  Bohannon}, {and} \bibinfo{person}{Fei Sha}.} \bibinfo{year}{2009}\natexlab{}.
\newblock \showarticletitle{{Robust web extraction: an approach based on a
  probabilistic tree-edit model}}. In \bibinfo{booktitle}{{\em International
  Conference on Management of Data}}.
\newblock
\showISBNx{9781605585512}
\showDOI{%
\url{https://doi.org/10.1145/1559845.1559882}}


\bibitem[\protect\citeauthoryear{Dalvi, Kumar, and Soliman}{Dalvi
  et~al\mbox{.}}{2011}]%
        {dalvi_automatic_2011}
\bibfield{author}{\bibinfo{person}{Nilesh Dalvi}, \bibinfo{person}{Ravi Kumar},
  {and} \bibinfo{person}{Mohamed Soliman}.} \bibinfo{year}{2011}\natexlab{}.
\newblock \showarticletitle{{Automatic wrappers for large scale web
  extraction}}.
\newblock \bibinfo{journal}{{\em Proceedings of the VLDB Endowment\/}}
  (\bibinfo{year}{2011}).
\newblock
\showISSN{2150-8097}


\bibitem[\protect\citeauthoryear{Fang, Xie, Zhang, Cheng, and Zhang}{Fang
  et~al\mbox{.}}{2017}]%
        {Fang2017}
\bibfield{author}{\bibinfo{person}{Yixiang Fang}, \bibinfo{person}{Xiaoqin
  Xie}, \bibinfo{person}{Xiaofeng Zhang}, \bibinfo{person}{Reynold Cheng},
  {and} \bibinfo{person}{Zhiqiang Zhang}.} \bibinfo{year}{2017}\natexlab{}.
\newblock \showarticletitle{{STEM: a suffix tree-based method for web data
  records extraction}}.
\newblock \bibinfo{journal}{{\em Knowledge and Information Systems\/}}
  (\bibinfo{year}{2017}).
\newblock
\showISSN{02193116}
\showDOI{%
\url{https://doi.org/10.1007/s10115-017-1062-0}}


\bibitem[\protect\citeauthoryear{Flesca, Manco, Masciari, Rende, and
  Tagarelli}{Flesca et~al\mbox{.}}{2004}]%
        {flesca_web_2004}
\bibfield{author}{\bibinfo{person}{S Flesca}, \bibinfo{person}{G Manco},
  \bibinfo{person}{E Masciari}, \bibinfo{person}{E Rende}, {and}
  \bibinfo{person}{A Tagarelli}.} \bibinfo{year}{2004}\natexlab{}.
\newblock \showarticletitle{{Web wrapper induction: a brief survey}}.
\newblock \bibinfo{journal}{{\em AI Communications\/}} (\bibinfo{year}{2004}).
\newblock


\bibitem[\protect\citeauthoryear{Gulhane, Madaan, Mehta, Ramamirtham, Rastogi,
  Satpal, Sengamedu, Tengli, and Tiwari}{Gulhane et~al\mbox{.}}{2011}]%
        {gulhane_web-scale_2011}
\bibfield{author}{\bibinfo{person}{P Gulhane}, \bibinfo{person}{A Madaan},
  \bibinfo{person}{R Mehta}, \bibinfo{person}{J Ramamirtham},
  \bibinfo{person}{R Rastogi}, \bibinfo{person}{S Satpal}, \bibinfo{person}{S~H
  Sengamedu}, \bibinfo{person}{A Tengli}, {and} \bibinfo{person}{C Tiwari}.}
  \bibinfo{year}{2011}\natexlab{}.
\newblock \showarticletitle{{Web-scale information extraction with vertex}}. In
  \bibinfo{booktitle}{{\em International Conference on Data Engineering}}.
  \bibinfo{publisher}{IEEE}.
\newblock
\showISBNx{978-1-4244-8959-6}
\showDOI{%
\url{https://doi.org/10.1109/ICDE.2011.5767842}}


\bibitem[\protect\citeauthoryear{Hao, Cai, Pang, and Zhang}{Hao
  et~al\mbox{.}}{2011}]%
        {hao_one_2011}
\bibfield{author}{\bibinfo{person}{Qiang Hao}, \bibinfo{person}{Rui Cai},
  \bibinfo{person}{Yanwei Pang}, {and} \bibinfo{person}{Lei Zhang}.}
  \bibinfo{year}{2011}\natexlab{}.
\newblock \showarticletitle{{From one tree to a forest: a unified solution for
  structured web data extraction}}. In \bibinfo{booktitle}{{\em International
  Conference on Research and Development in Information Retrieval}}.
\newblock


\bibitem[\protect\citeauthoryear{Jindal and Liu}{Jindal and Liu}{2010}]%
        {jindal_generalized_2010}
\bibfield{author}{\bibinfo{person}{Nitin Jindal} {and} \bibinfo{person}{Bing
  Liu}.} \bibinfo{year}{2010}\natexlab{}.
\newblock \showarticletitle{{A Generalized Tree Matching Algorithm Considering
  Nested Lists for Web Data Extraction}}.
\newblock \bibinfo{journal}{{\em The SIAM International Conference on Data
  Mining\/}} (\bibinfo{year}{2010}).
\newblock
\showISBNx{9781611972801}
\showDOI{%
\url{https://doi.org/10.1137/1.9781611972801.81}}


\bibitem[\protect\citeauthoryear{Kayed, Kayed, Girgis, and Shaalan}{Kayed
  et~al\mbox{.}}{2006}]%
        {kayed_survey_2006}
\bibfield{author}{\bibinfo{person}{M Kayed}, \bibinfo{person}{M Kayed},
  \bibinfo{person}{M~R Girgis}, {and} \bibinfo{person}{K~F Shaalan}.}
  \bibinfo{year}{2006}\natexlab{}.
\newblock \showarticletitle{{A survey of web information extraction systems}}.
\newblock \bibinfo{journal}{{\em IEEE Transactions on Knowledge and Data
  Engineering\/}} (\bibinfo{year}{2006}).
\newblock


\bibitem[\protect\citeauthoryear{Kosala, {Van den Bussche}, Bruynooghe, and
  Blockeel}{Kosala et~al\mbox{.}}{2002}]%
        {Kosala2002}
\bibfield{author}{\bibinfo{person}{Raymond Kosala}, \bibinfo{person}{Jan {Van
  den Bussche}}, \bibinfo{person}{Maurice Bruynooghe}, {and}
  \bibinfo{person}{Hendrik Blockeel}.} \bibinfo{year}{2002}\natexlab{}.
\newblock \showarticletitle{{Information Extraction in Structured Documents
  Using Tree Automata Induction}}.
\newblock \bibinfo{publisher}{Springer, Berlin, Heidelberg}.
\newblock
\showDOI{%
\url{https://doi.org/10.1007/3-540-45681-3_25}}


\bibitem[\protect\citeauthoryear{Kushmerick}{Kushmerick}{2000}]%
        {kushmerick_wrapper_2000}
\bibfield{author}{\bibinfo{person}{N Kushmerick}.}
  \bibinfo{year}{2000}\natexlab{}.
\newblock \showarticletitle{{Wrapper induction: Efficiency and
  expressiveness}}.
\newblock \bibinfo{journal}{{\em Artificial Intelligence\/}}
  (\bibinfo{year}{2000}).
\newblock


\bibitem[\protect\citeauthoryear{Kushmerick, Weld, and Doorenbos}{Kushmerick
  et~al\mbox{.}}{1997}]%
        {kushmerick_wrapper_1997}
\bibfield{author}{\bibinfo{person}{Nicholas Kushmerick},
  \bibinfo{person}{Daniel~S. Weld}, {and} \bibinfo{person}{Robert Doorenbos}.}
  \bibinfo{year}{1997}\natexlab{}.
\newblock \bibinfo{title}{{Wrapper induction for information extraction}}.
\newblock   (\bibinfo{year}{1997}).
\newblock
\showISBNx{0591708434}
\showDOI{%
\url{https://doi.org/10.1.1.33.2176}}


\bibitem[\protect\citeauthoryear{Nasti, Asghar, and Butt}{Nasti
  et~al\mbox{.}}{2016}]%
        {Nasti2016}
\bibfield{author}{\bibinfo{person}{Samiah~Jan Nasti}, \bibinfo{person}{M.
  Asghar}, {and} \bibinfo{person}{Muheet~Ahmad Butt}.}
  \bibinfo{year}{2016}\natexlab{}.
\newblock \showarticletitle{{A Comparative Study on Web Data Extraction
  Approaches}}.
\newblock \bibinfo{journal}{{\em International Journal of Engineering Science
  and Computing\/}} (\bibinfo{year}{2016}).
\newblock
\showDOI{%
\url{https://doi.org/10.4010/2016.1322}}


\bibitem[\protect\citeauthoryear{Omari, Shoham, and Yahav}{Omari
  et~al\mbox{.}}{2017}]%
        {Omari2017}
\bibfield{author}{\bibinfo{person}{Adi Omari}, \bibinfo{person}{Sharon Shoham},
  {and} \bibinfo{person}{Eran Yahav}.} \bibinfo{year}{2017}\natexlab{}.
\newblock \showarticletitle{{Synthesis of Forgiving Data Extractors}}.
\newblock \bibinfo{journal}{{\em International Conference on Web Search and
  Data Mining\/}} (\bibinfo{year}{2017}).
\newblock
\showISBNx{9781450346757}
\showDOI{%
\url{https://doi.org/10.1145/3018661.3018740}}


\bibitem[\protect\citeauthoryear{Parameswaran, Dalvi, Garcia-Molina, and
  Rastogi}{Parameswaran et~al\mbox{.}}{2011}]%
        {parameswaran_optimal_2011}
\bibfield{author}{\bibinfo{person}{Aditya Parameswaran}, \bibinfo{person}{N
  Dalvi}, \bibinfo{person}{H Garcia-Molina}, {and} \bibinfo{person}{R
  Rastogi}.} \bibinfo{year}{2011}\natexlab{}.
\newblock \showarticletitle{{Optimal Schemes for Robust Web Extraction}}. In
  \bibinfo{booktitle}{{\em International Conference on Very Large Data Bases}}.
\newblock


\bibitem[\protect\citeauthoryear{Reis, Golgher, Silva, and Laender}{Reis
  et~al\mbox{.}}{2004}]%
        {reis_automatic_2004}
\bibfield{author}{\bibinfo{person}{D~C Reis}, \bibinfo{person}{Paulo~B
  Golgher}, \bibinfo{person}{A~S Silva}, {and} \bibinfo{person}{a~F Laender}.}
  \bibinfo{year}{2004}\natexlab{}.
\newblock \showarticletitle{{Automatic web news extraction using tree edit
  distance}}. In \bibinfo{booktitle}{{\em World Wide Web}}.
\newblock
\showISBNx{1-58113-844-X}
\showDOI{%
\url{https://doi.org/10.1145/988672.988740}}


\bibitem[\protect\citeauthoryear{Yang}{Yang}{1991}]%
        {yang_identifying_1991}
\bibfield{author}{\bibinfo{person}{W Yang}.} \bibinfo{year}{1991}\natexlab{}.
\newblock \showarticletitle{{Identifying syntactic differences between two
  programs}}.
\newblock \bibinfo{journal}{{\em Software - Practice and Experience\/}}
  (\bibinfo{year}{1991}).
\newblock


\bibitem[\protect\citeauthoryear{Zhai and Liu}{Zhai and Liu}{2005}]%
        {zhai_web_2005}
\bibfield{author}{\bibinfo{person}{Y Zhai} {and} \bibinfo{person}{B Liu}.}
  \bibinfo{year}{2005}\natexlab{}.
\newblock \showarticletitle{{Web data extraction based on partial tree
  alignment}}. In \bibinfo{booktitle}{{\em World Wide Web}}.
\newblock


\bibitem[\protect\citeauthoryear{Zheng, Song, Wen, and Giles}{Zheng
  et~al\mbox{.}}{2009}]%
        {zheng_efficient_2009}
\bibfield{author}{\bibinfo{person}{S Zheng}, \bibinfo{person}{R Song},
  \bibinfo{person}{J~R Wen}, {and} \bibinfo{person}{C~L Giles}.}
  \bibinfo{year}{2009}\natexlab{}.
\newblock \showarticletitle{{Efficient record-level wrapper induction}}. In
  \bibinfo{booktitle}{{\em Conference on Information and knowledge
  Management}}.
\newblock


\end{thebibliography}

\end{document}